\documentclass[a4paper,11pt]{article}
\pdfoutput=1 

\usepackage{jheppub} 

\usepackage[T1]{fontenc} 

\usepackage{graphicx,color,dcolumn,booktabs,bm}
\usepackage{longtable,lscape}
\usepackage{amsmath}
\usepackage{indentfirst}
\usepackage{epsfig}
\usepackage{epstopdf}
\usepackage{slashed}
\usepackage{color}
\usepackage{multirow}
\usepackage{graphicx,color,dcolumn,booktabs,bm}
\usepackage{verbatim}
\usepackage{mathrsfs}
\usepackage{rotating}
\usepackage{makecell}
\usepackage{threeparttable}
\usepackage{enumerate}
\usepackage{subfigure}
\usepackage{float}
\usepackage{setspace}
\usepackage{lineno}
\newcommand{\ra}{\rightarrow}
\newcommand{\EE}{e^+e^-}
\newcommand{\lamc}{\Lambda_c^+}

\newcommand{\pkpi}{pK^-\pi^+}

\newcommand{\petap}{p\eta'}

\newcommand{\ppipieta}{p\pi^{+}\pi^{-}\eta}
\newcommand{\GG}{\gamma\gamma}
\newcommand{\BF}{\mathcal{B}}

\title{\boldmath First Measurement of the $\Lambda_c^+ \to p \eta'$ decay}

\newcounter{AffiliationCounter}
\stepcounter{AffiliationCounter}\edef\instBilbao{\protect\theAffiliationCounter}
\stepcounter{AffiliationCounter}\edef\instBonn{\protect\theAffiliationCounter}
\stepcounter{AffiliationCounter}\edef\instBNL{\protect\theAffiliationCounter}
\stepcounter{AffiliationCounter}\edef\instBINP{\protect\theAffiliationCounter}
\stepcounter{AffiliationCounter}\edef\instCharles{\protect\theAffiliationCounter}
\stepcounter{AffiliationCounter}\edef\instChonnam{\protect\theAffiliationCounter}
\stepcounter{AffiliationCounter}\edef\instCAU{\protect\theAffiliationCounter}
\stepcounter{AffiliationCounter}\edef\instCincinnati{\protect\theAffiliationCounter}
\stepcounter{AffiliationCounter}\edef\instDESY{\protect\theAffiliationCounter}
\stepcounter{AffiliationCounter}\edef\instDuke{\protect\theAffiliationCounter}
\stepcounter{AffiliationCounter}\edef\instDuyTan{\protect\theAffiliationCounter}
\stepcounter{AffiliationCounter}\edef\instFlorida{\protect\theAffiliationCounter}
\stepcounter{AffiliationCounter}\edef\instFuJen{\protect\theAffiliationCounter}
\stepcounter{AffiliationCounter}\edef\instFudan{\protect\theAffiliationCounter}
\stepcounter{AffiliationCounter}\edef\instGifu{\protect\theAffiliationCounter}
\stepcounter{AffiliationCounter}\edef\instSokendai{\protect\theAffiliationCounter}
\stepcounter{AffiliationCounter}\edef\instGyeongsang{\protect\theAffiliationCounter}
\stepcounter{AffiliationCounter}\edef\instHanyang{\protect\theAffiliationCounter}
\stepcounter{AffiliationCounter}\edef\instHawaii{\protect\theAffiliationCounter}
\stepcounter{AffiliationCounter}\edef\instKEK{\protect\theAffiliationCounter}
\stepcounter{AffiliationCounter}\edef\instJPARC{\protect\theAffiliationCounter}
\stepcounter{AffiliationCounter}\edef\instHSE{\protect\theAffiliationCounter}
\stepcounter{AffiliationCounter}\edef\instIKER{\protect\theAffiliationCounter}
\stepcounter{AffiliationCounter}\edef\instIISERM{\protect\theAffiliationCounter}
\stepcounter{AffiliationCounter}\edef\instIITG{\protect\theAffiliationCounter}
\stepcounter{AffiliationCounter}\edef\instIITH{\protect\theAffiliationCounter}
\stepcounter{AffiliationCounter}\edef\instIITM{\protect\theAffiliationCounter}
\stepcounter{AffiliationCounter}\edef\instIndiana{\protect\theAffiliationCounter}
\stepcounter{AffiliationCounter}\edef\instProtvino{\protect\theAffiliationCounter}
\stepcounter{AffiliationCounter}\edef\instVienna{\protect\theAffiliationCounter}
\stepcounter{AffiliationCounter}\edef\instNapoli{\protect\theAffiliationCounter}
\stepcounter{AffiliationCounter}\edef\instRomaTre{\protect\theAffiliationCounter}
\stepcounter{AffiliationCounter}\edef\instTorino{\protect\theAffiliationCounter}
\stepcounter{AffiliationCounter}\edef\instISU{\protect\theAffiliationCounter}
\stepcounter{AffiliationCounter}\edef\instJAEA{\protect\theAffiliationCounter}
\stepcounter{AffiliationCounter}\edef\instJSI{\protect\theAffiliationCounter}
\stepcounter{AffiliationCounter}\edef\instKarlsruhe{\protect\theAffiliationCounter}
\stepcounter{AffiliationCounter}\edef\instIPMU{\protect\theAffiliationCounter}
\stepcounter{AffiliationCounter}\edef\instKAU{\protect\theAffiliationCounter}
\stepcounter{AffiliationCounter}\edef\instKitasato{\protect\theAffiliationCounter}
\stepcounter{AffiliationCounter}\edef\instKISTI{\protect\theAffiliationCounter}
\stepcounter{AffiliationCounter}\edef\instKorea{\protect\theAffiliationCounter}
\stepcounter{AffiliationCounter}\edef\instKyungpook{\protect\theAffiliationCounter}
\stepcounter{AffiliationCounter}\edef\instIJCLab{\protect\theAffiliationCounter}
\stepcounter{AffiliationCounter}\edef\instLebedev{\protect\theAffiliationCounter}
\stepcounter{AffiliationCounter}\edef\instLjubljana{\protect\theAffiliationCounter}
\stepcounter{AffiliationCounter}\edef\instLuther{\protect\theAffiliationCounter}
\stepcounter{AffiliationCounter}\edef\instMNIT{\protect\theAffiliationCounter}
\stepcounter{AffiliationCounter}\edef\instMaribor{\protect\theAffiliationCounter}
\stepcounter{AffiliationCounter}\edef\instMPI{\protect\theAffiliationCounter}
\stepcounter{AffiliationCounter}\edef\instMelbourne{\protect\theAffiliationCounter}
\stepcounter{AffiliationCounter}\edef\instMississippi{\protect\theAffiliationCounter}
\stepcounter{AffiliationCounter}\edef\instMiyazaki{\protect\theAffiliationCounter}
\stepcounter{AffiliationCounter}\edef\instMEPhI{\protect\theAffiliationCounter}
\stepcounter{AffiliationCounter}\edef\instNagoya{\protect\theAffiliationCounter}
\stepcounter{AffiliationCounter}\edef\instUNapoli{\protect\theAffiliationCounter}
\stepcounter{AffiliationCounter}\edef\instNara{\protect\theAffiliationCounter}
\stepcounter{AffiliationCounter}\edef\instNCU{\protect\theAffiliationCounter}
\stepcounter{AffiliationCounter}\edef\instTaiwan{\protect\theAffiliationCounter}
\stepcounter{AffiliationCounter}\edef\instKrakow{\protect\theAffiliationCounter}
\stepcounter{AffiliationCounter}\edef\instNihonDental{\protect\theAffiliationCounter}
\stepcounter{AffiliationCounter}\edef\instNiigata{\protect\theAffiliationCounter}
\stepcounter{AffiliationCounter}\edef\instNovaGorica{\protect\theAffiliationCounter}
\stepcounter{AffiliationCounter}\edef\instNovosibirsk{\protect\theAffiliationCounter}
\stepcounter{AffiliationCounter}\edef\instOsakaCity{\protect\theAffiliationCounter}
\stepcounter{AffiliationCounter}\edef\instPNNL{\protect\theAffiliationCounter}
\stepcounter{AffiliationCounter}\edef\instPanjab{\protect\theAffiliationCounter}
\stepcounter{AffiliationCounter}\edef\instPittsburgh{\protect\theAffiliationCounter}
\stepcounter{AffiliationCounter}\edef\instPunjab{\protect\theAffiliationCounter}
\stepcounter{AffiliationCounter}\edef\instNPC{\protect\theAffiliationCounter}
\stepcounter{AffiliationCounter}\edef\instRIKENMSL{\protect\theAffiliationCounter}
\stepcounter{AffiliationCounter}\edef\instURomaTre{\protect\theAffiliationCounter}
\stepcounter{AffiliationCounter}\edef\instUSTC{\protect\theAffiliationCounter}
\stepcounter{AffiliationCounter}\edef\instShoyaku{\protect\theAffiliationCounter}
\stepcounter{AffiliationCounter}\edef\instSoochow{\protect\theAffiliationCounter}
\stepcounter{AffiliationCounter}\edef\instSoongsil{\protect\theAffiliationCounter}
\stepcounter{AffiliationCounter}\edef\instSungkyunkwan{\protect\theAffiliationCounter}
\stepcounter{AffiliationCounter}\edef\instSydney{\protect\theAffiliationCounter}
\stepcounter{AffiliationCounter}\edef\instTabuk{\protect\theAffiliationCounter}
\stepcounter{AffiliationCounter}\edef\instTata{\protect\theAffiliationCounter}
\stepcounter{AffiliationCounter}\edef\instToho{\protect\theAffiliationCounter}
\stepcounter{AffiliationCounter}\edef\instTohoku{\protect\theAffiliationCounter}
\stepcounter{AffiliationCounter}\edef\instERI{\protect\theAffiliationCounter}
\stepcounter{AffiliationCounter}\edef\instTokyo{\protect\theAffiliationCounter}
\stepcounter{AffiliationCounter}\edef\instTIT{\protect\theAffiliationCounter}
\stepcounter{AffiliationCounter}\edef\instTMU{\protect\theAffiliationCounter}
\stepcounter{AffiliationCounter}\edef\instVPI{\protect\theAffiliationCounter}
\stepcounter{AffiliationCounter}\edef\instWayneState{\protect\theAffiliationCounter}
\stepcounter{AffiliationCounter}\edef\instYamagata{\protect\theAffiliationCounter}
\stepcounter{AffiliationCounter}\edef\instYonsei{\protect\theAffiliationCounter}

\collaboration{The Belle Collaboration}
  \author[\instFudan]{S.~X.~Li,} 
  \author[\instFudan]{J.~X.~Cui,} 
  \author[\instFudan]{C.~P.~Shen,} 
  \author[\instKEK,\instSokendai]{I.~Adachi,} 
  \author[\instTokyo]{H.~Aihara,} 
  \author[\instTabuk,\instKAU]{S.~Al~Said,} 
  \author[\instBNL]{D.~M.~Asner,} 
  \author[\instCincinnati]{H.~Atmacan,} 
  \author[\instHSE]{T.~Aushev,} 
  \author[\instTabuk]{R.~Ayad,} 
% ...

\abstract{
     We present the first measurement of the branching fraction of the singly Cabibbo-suppressed (SCS) decay $\Lambda_c^+ \ra p \eta'$ with $\eta' \ra \eta\pi^+\pi^-$, using a data sample corresponding to an integrated luminosity of 981 $\rm fb^{-1}$, collected by the Belle detector at the KEKB $e^{+}$$e^{-}$ asymmetric-energy collider.  A significant $\Lambda_c^+ \ra p\eta'$ signal is observed for the first time with a signal significance of 5.4$\sigma$. The relative branching fraction with respect to the normalization mode $\Lambda_c^+ \ra pK^-\pi^+$ is measured to be
     \begin{equation*}
     \frac{{\cal B}(\Lambda_c^+ \ra p \eta')}{{\cal B}(\Lambda_c^+ \ra p K^-\pi^+)} = (7.54 \pm 1.32  \pm 0.73) \times 10^{-3},
     \end{equation*}
     where the uncertainties are statistical and systematic, respectively. Using the world-average value of ${\cal B}(\Lambda_c^+ \ra pK^-\pi^+) = (6.28\pm0.32)\times10^{-2}$, we obtain
     \begin{equation*}
     {\cal B}(\Lambda_c^+ \ra p \eta') = (4.73 \pm 0.82 \pm 0.46 \pm 0.24)\times 10^{-4},
     \end{equation*}
     where the uncertainties are statistical, systematic, and from ${\cal B}(\Lambda_c^+ \ra pK^-\pi^+)$, respectively.
}

\begin{document}
\maketitle
\flushbottom

\section{Introduction}
\label{sec:intro}
     Hadronic decays of charmed baryons provide an ideal laboratory to understand the interplay of weak and strong interactions in the charm system~\cite{lamcR1,lamcR2,lamcR3}. Decays of charmed baryons receive sizable nonfactorizable contributions from W-exchange diagrams, which are subject to color and helicity suppression~\cite{lamcR4,lamcR5,lamcR6,lamcR7}. Therefore, the study of nonfactorizable contributions is critical to understand the dynamics of charmed baryon decays. To avoid theoretical difficulties in the factorization approach~\cite{lamcR4}, one can use the $SU(3)_F$ flavor symmetry to relate the amplitudes among different decays~\cite{lamcR5,lamcR8,lamcR9}. Other theoretical approaches provide calculations based on dynamical models~\cite{lamcR11,lamcR12,lamcR13}. For the singly Cabibbo-suppressed (SCS) decay $\lamc \ra \petap$, theoretical predictions on its branching fraction under different assumptions vary by more than an order of magnitude as listed in table~\ref{theory-bf}. Currently, this decay has not yet been observed.
     \begin{table}[H]
             \centering
             \caption{Comparison of different theoretical predictions for $\BF(\lamc \ra \petap)$ (in units of $10^{-3}$).}
             \label{theory-bf}
             \small
             \setlength{\tabcolsep}{1.5mm}{
             \begin{tabular}{c | c | c | c }
             \hline \hline
               &$SU(3)_F$ symmetry~\cite{lamcR5} &$SU(3)_F$ symmetry~\cite{Ta3} & Constituent quark model~\cite{lamcR3}        \\ \hline
        $\BF(\lamc \ra \petap)$  & $0.4-0.6$                 &$1.22^{+1.43}_{-0.87}$    & $0.04-0.2$   \\ \hline \hline
             \end{tabular} }
     \end{table}

     In this study, based on an $e^+e^-$ annihilation data sample of 981 $\rm fb^{-1}$ collected by the Belle experiment, we measure the branching fraction of the signal mode $\lamc \ra \petap$ with respect to the normalization mode $\lamc \ra \pkpi$. Throughout this paper, charge-conjugate modes are implicitly included unless stated otherwise.
     The paper is organized as follows. Section~\ref{sec:dataset} introduces the Belle detector and data sample. Section~\ref{sec:sel} discusses the event selection criteria. The signal and background estimations are presented in section~\ref{sec:fit}. Sections~\ref{sec:syserror} and \ref{sec:conc} describe the systematic uncertainty and conclusion, respectively.

\section{The Belle detector and data sample}
\label{sec:dataset}
     This measurement is based on a data sample corresponding to an integrated luminosity of 981 $\rm fb^{-1}$, collected with the Belle detector at the KEKB asymmetric-energy $e^+e^-$ collider~\cite{KEKB1,KEKB2}. About 70\% of the data were recorded at the $\Upsilon(4S)$ resonance, and the rest were collected at other $\Upsilon(nS)$ ($n$ = 1,~2,~3, or 5) states or at center-of-mass (CM) energies a few tens of MeV below the $\Upsilon(4S)$ or the $\Upsilon(nS)$ peaks.

     The Belle detector is a large-solid-angle magnetic spectrometer that consists of a silicon vertex detector (SVD), a 50-layer central drift chamber (CDC), an array of aerogel threshold Cherenkov counters (ACC), a barrel-like arrangement of time-of-flight scintillation counters (TOF), and an electromagnetic calorimeter comprised of CsI(Tl) crystals (ECL) located inside a superconducting solenoid coil that provides a 1.5~T magnetic field. An iron flux-return located outside of the coil is instrumented to detect $K_L^0$ mesons and to identify muons. The detector is described in detail elsewhere~\cite{Belle1,Belle2}. The origin of the coordinate system is defined as the position of the nominal interaction point, and the axis aligning with the direction opposite the $e^+$ beam is defined as the $z$ axis.

     Monte Carlo (MC) simulated events are used to optimize the selection criteria, study backgrounds, and determine the signal reconstruction efficiency. Samples of simulated signal MC events are generated by {\sc EvtGen}~\cite{evtgen} and propagated through a detector simulation based on {\sc geant3}~\cite{geant3}. The $e^+e^- \ra c\bar{c}$ events are simulated using {\sc pythia}~\cite{pythia}; the decays $\lamc \ra \pkpi$ and $\eta' \ra \eta\pi^+\pi^-$ are generated with a phase space model. We take into account the effect of final-state radiation from charged particles by using the {\sc photos} package~\cite{photons}. Simulated samples of $\Upsilon(4S)\ra B^{+}B^{-}/B^{0}\bar{B}^{0}$, $\Upsilon(5S)\ra B_{s}^{(*)}\bar{B}_{s}^{(*)}/B^{(*)}\bar{B}^{(*)}(\pi)/\Upsilon(4S)\gamma$, $\EE \to q\bar{q}$ $(q=u,~d,~s,~c)$ at $\sqrt{s}$ = 10.52, 10.58, and 10.867~GeV, and $\Upsilon(1S,~2S,~3S)$ decays, normalized to the same integrated luminosity as real data, are used to develop the selection criteria and perform the background study~\cite{topoana}.

\section{Selection criteria}
\label{sec:sel}
     Selection criteria are optimized by maximizing a figure-of-merit $\epsilon/(\frac{a}{2}+\sqrt{n_{\rm B}})$~\cite{fomcite}, where $\epsilon$ is the signal efficiency; $\it a$ is the target signal significance expressed in standard deviations in a one-sided Gaussian test, selected to be 5; $n_{\rm B}$ is the number of background events expected in a two-dimensional signal region of $\eta'$ and $\lamc$ signals, which is defined as (0.95, 0.965) GeV/$c^2$ in $M(\eta\pi^+\pi^-)$ and (2.27, 2.31) GeV/$c^2$ in $M(\petap)$.

     We reconstruct the decays $\lamc \ra \petap$ and $\lamc \ra \pkpi$, with the $\eta'$ decay reconstructed in the cascade $\eta\pi^+\pi^-$, $\eta \ra \gamma\gamma$. Final-state charged tracks are identified as $p$, $K$, or $\pi$ candidates using information from the charged-hadron identification systems (ACC, TOF, CDC) combined into a likelihood ratio, $\mathcal{R}(h|h') = \mathcal{L}(h)/(\mathcal{L}(h)+\mathcal{L}(h'))$ where $h$ and $h'$ are $\pi$, $K$, or $p$ as appropriate~\cite{pidcode}. Tracks having $\mathcal{R}(p|\pi)>0.9$ and $\mathcal{R}(p|K)>0.9$ are identified as proton candidates; charged kaon candidates are required to have $\mathcal{R}(K|p) > 0.4$ and $\mathcal{R}(K|\pi) > 0.9$; and charged pion candidates to have $\mathcal{R}(\pi|p) > 0.4$ and $\mathcal{R}(\pi|K) > 0.4$. A likelihood ratio for electron identification, $\mathcal{R}(e)$, is formed from ACC, CDC, and ECL information~\cite{eidcode}, and is required to be less than 0.9 for all charged tracks to suppress electrons.
     The identification efficiencies of $p$, $K$, and $\pi$ are 82\%, 70\%, and 97\%, respectively. The probabilities of misidentifying $h$ as $h'$, $P(h\ra h')$, are estimated to be 3\% [$P(p\ra \pi)$], 7\% [$P(p\ra K)$], 10\% [$P(K\ra \pi)$], 2\% [$P(K\ra p)$], 5\% [$P(\pi\ra K)$], and 1\% [$P(\pi\ra p)$]. For each charged track, the distance of the closest approach with respect to the interaction point along the $z$ axis and in the transverse $x-y$ plane is required to be less than 2.0 cm and 0.1 cm, respectively. Each track must have at least one SVD hit in both the $z$ direction and the $x-y$ plane.

     Photon candidates are selected from ECL clusters not associated with any charged tracks. The photon energy is required to be greater than 90 MeV in the barrel region ($-0.63<$ cos$\theta$ $<0.85$) and greater than 120 MeV in the endcap region ($-0.91<$ cos$\theta < -0.63$ or $0.85<$ cos$\theta<0.98$) of the ECL, where $\theta$ is the polar angle relative to the positive $z$ axis. To reject neutral hadrons, the ratio of the energy deposited in the central $3\times3$ array of ECL crystals to the total energy deposited in the enclosing $5\times5$ array of crystals is required to be at least 0.9 for each photon candidate.

     The $\eta$ candidates are reconstructed via their decay to two photons. The $\gamma\gamma$ invariant mass is required to satisfy $0.45 < M(\gamma\gamma) < 0.65$ GeV/$c^2$, and then a mass-constrained fit is performed for $\eta$ candidates to improve the momentum resolution. The corresponding $\chi^2$ value of the mass-constrained fit on $\eta$ ($\chi^2_{\eta}$) is required to be less than 10.
     To further suppress background events, we remove $\eta$ candidates in which either of the daughter photons can be combined with other photons in the event to form $\pi^0 \ra \gamma\gamma$ candidates satisfying $|M(\gamma\gamma)-m_{\pi^0}|<< 12$ MeV/$c^2$, where $m_{\pi^0}$ is the nominal $\pi^0$ mass~\cite{pdg}. With this veto, we reject 42\% of the background, while retaining 83\% of the signal.

\begin{figure*}[h!tbp]
          \begin{center}
          \includegraphics[width=7cm]{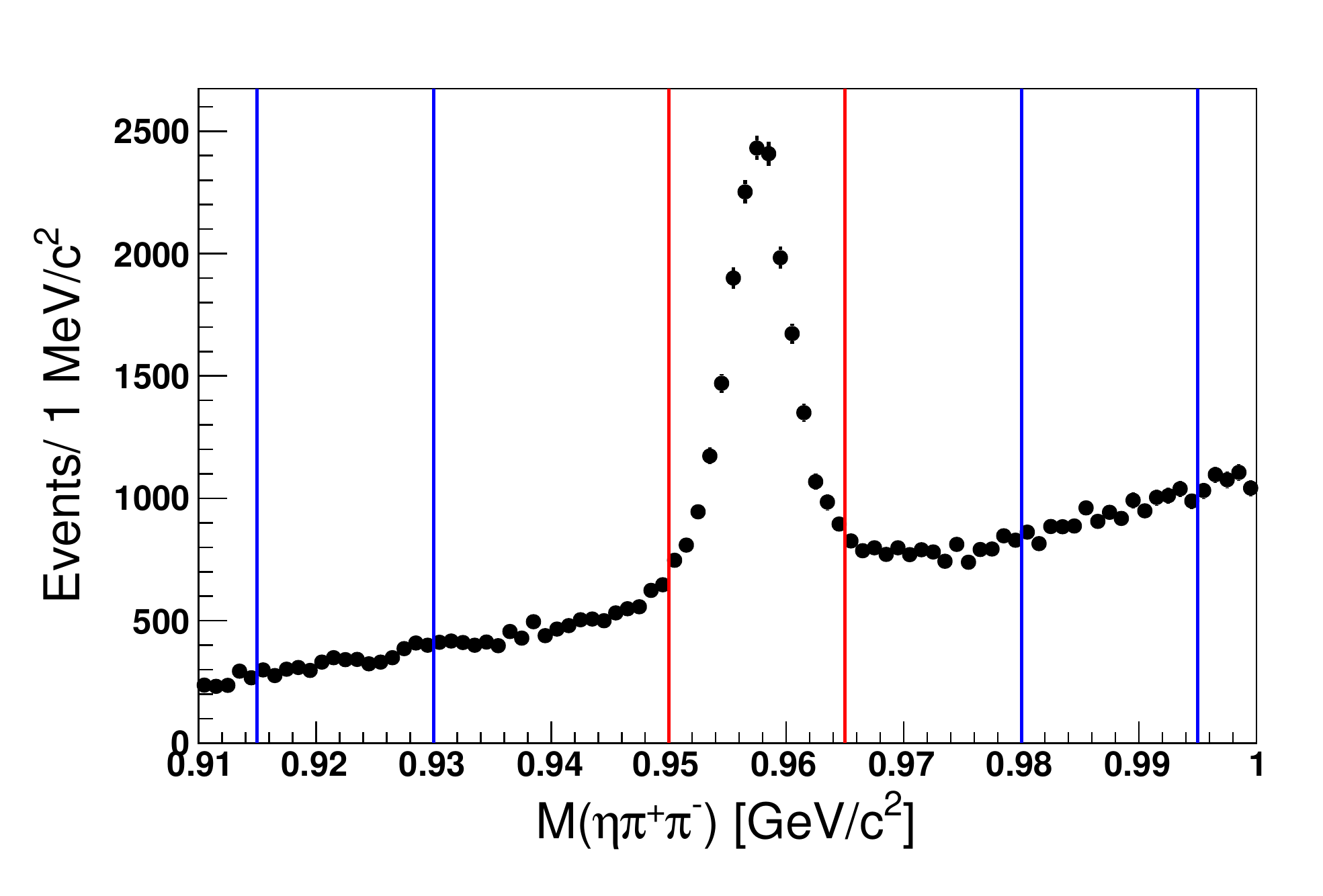}
          \caption{\label{invm-etap} The invariant mass distribution of $\eta\pi^+\pi^-$. The region between two red lines is selected as the $\eta'$ signal region, and the regions between two blue lines are the $\eta'$ sidebands.}
          \end{center}
     \end{figure*}

     The $\eta'$ candidates are reconstructed by combining two opposite-charge $\pi$ tracks with an $\eta$ candidate. The invariant mass distribution of $\eta\pi^+\pi^-$ from data is shown in figure~\ref{invm-etap}. Candidates $\eta'$ are retained if $0.950 < M(\eta\pi^+\pi^-) < 0.965$ GeV/$c^2$, corresponding to an efficiency of 96\%. The $\eta'$ sidebands are defined as 0.915 to 0.930 GeV/$c^2$ and 0.980 to 0.995 GeV/$c^2$, which are the regions between two blue lines in the $M(\eta\pi^+\pi^-)$ distribution.

     Candidates for $\lamc \ra \pkpi$ and $\lamc \ra \petap$ decays are reconstructed by combining $p$, $K^-$, $\pi^+$ candidates, and $p$, $\eta'$ candidates, respectively. A vertex fit is performed with the three charged tracks to suppress combinatorial background events. The resulting fit quality is labeled $\chi^2_{\rm vtx}$. For $\lamc \ra \pkpi$, the $\chi^2_{\rm vtx}$ is required to be less than 40, while for $\lamc \ra \petap$, $\chi^2_{\rm vtx} \textless 15$ is required. For both $\chi^2_{\rm vtx}$ requirements, the efficiency is larger than 98\%. A scaled momentum of $x_p > 0.53$ is required to suppress background, especially from $B$-meson decays, where $x_{p} = {p^{*}}/{\sqrt{E^{2}_{\rm cm}/4c^2 - M^2c^2}}$, $E_{\rm cm}$ is the CM energy, and $p^{*}$ and $M$ are the momentum and invariant mass, respectively, of the $\Lambda_c^+$ candidates in the CM frame.

     After the preliminary selection, about 0.8\% of the $\lamc \ra \pkpi$ events and 13.3\% of the $\lamc \ra \petap$ events have two or more $\lamc$ candidates. We choose the best $\eta$ candidate according to the smallest value of $\chi^2_{\eta}$; the rate of events having multiple $\lamc \ra \petap$ candidates with this criterion is 1.6\%. For such multi-candidate events, we choose a single $\lamc$ candidate randomly. This best-candidate selection, based on the MC simulation, identifies the correct candidate 65\% of the time. The $\petap$ mass distribution for wrong-combination simulated signal events is found to be smooth.  We keep all candidates of $\lamc \ra \pkpi$ in multiple-candidate events as the multiplicity is negligible.

\section{Signal and background estimation}
\label{sec:fit}
\begin{figure*}[h!tbp]
          \begin{center}
          \includegraphics[width=7cm]{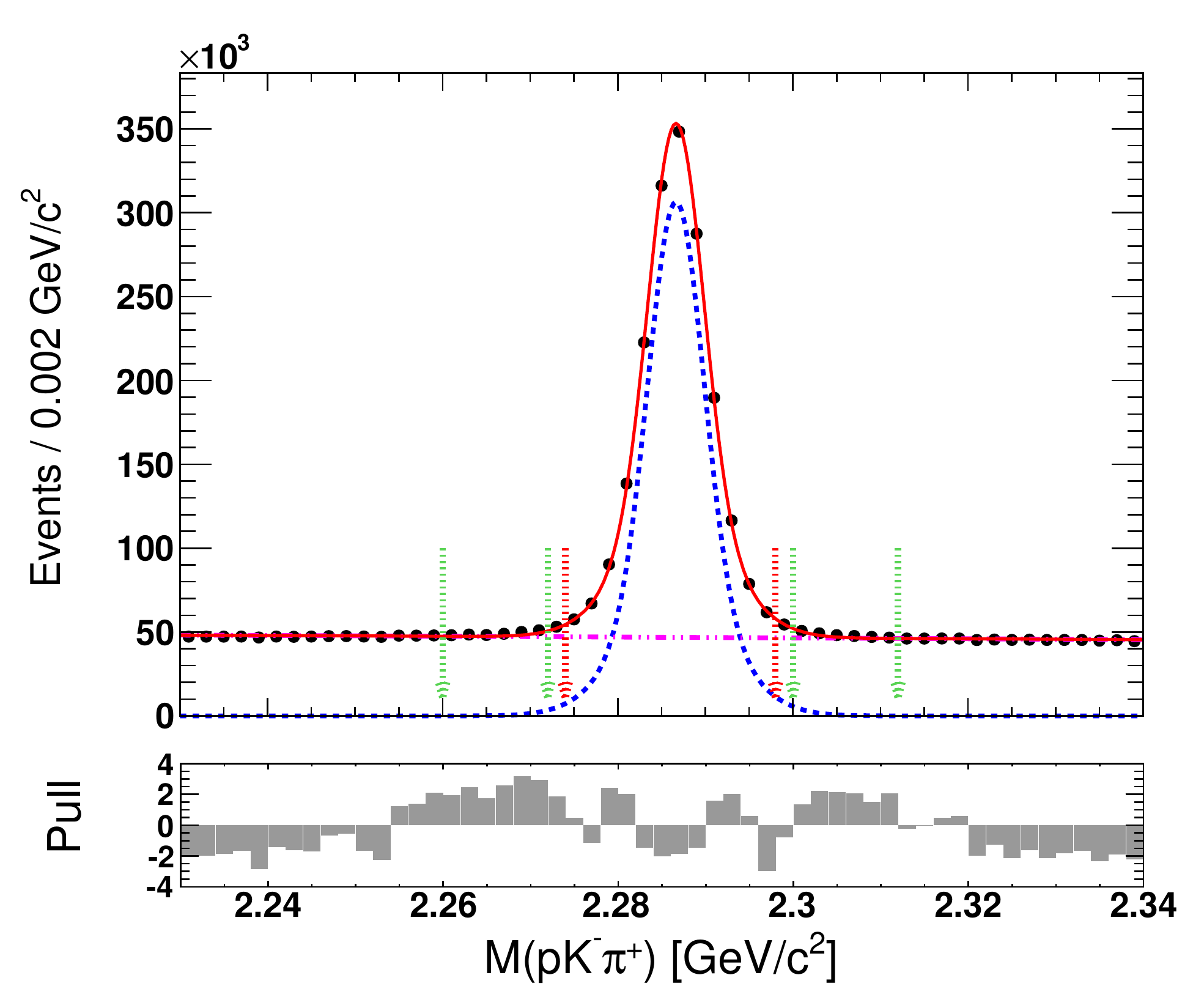}
          \includegraphics[width=7cm]{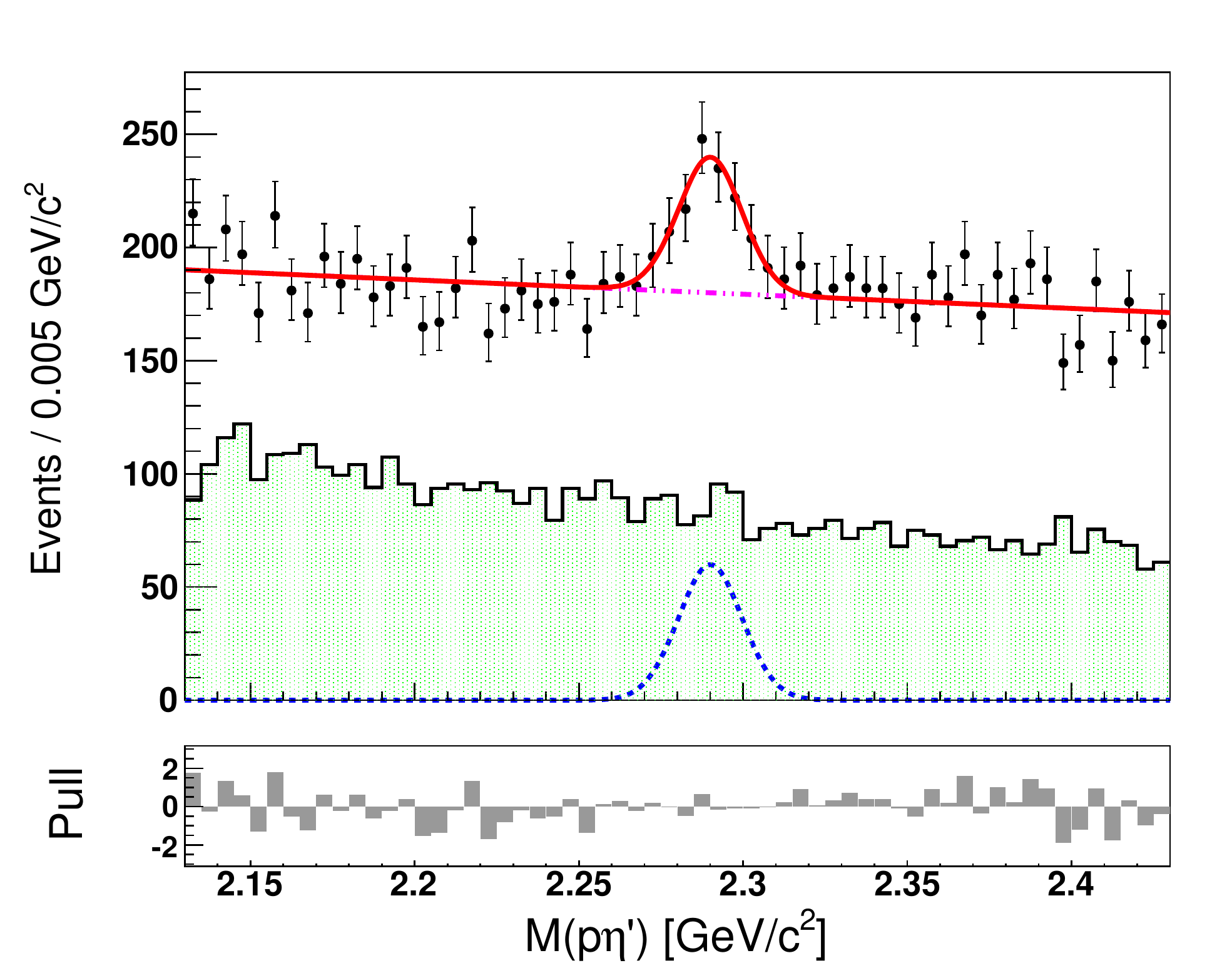}
          \caption{\label{lamc-data-invM-fit} Fits to the invariant mass distributions of the $\pkpi$ (Left) and $\petap$ combinations (Right). Black dots with error bars represent the data; red solid lines represent the total fitted result; blue dashed lines represent the signal shape; magenta dot-dashed lines represent the background shape; and the green histogram is from normalized $\eta'$ sidebands.}
          \end{center}
     \end{figure*}

     With the above selection criteria applied, the invariant mass distributions of normalization and signal modes are shown in figure~\ref{lamc-data-invM-fit}.
     From a study of generic MC samples~\cite{topoana}, no known peaking background processes contribute to mass distributions in the $\lamc$ signal region.

     To extract the number of signal events, we perform an unbinned maximum-likelihood fit to the $M(\pkpi)$ or $M(p\eta')$ distribution. The likelihood function is defined in terms of a signal PDF ($F_{\rm S}$) and a background PDF ($F_{\rm B}$) as
     \begin{equation}
           \label{eq:likelihood}
           \mathcal{L}=\frac{e^{-(n_{\rm S}+n_{\rm B})}}{N!}\prod_i^{N} \left[n_{\rm S}F_{\rm S}(M_i)+n_{\rm B}F_{\rm B}(M_i)\right],
     \end{equation}
     where $N$ is the total number of observed events; $n_{\rm S}$ and $n_{\rm B}$ are the numbers of signal events and background events, respectively; $M$ is $\pkpi$ or $\petap$ invariant mass; and $i$ denotes the event index. The fit is performed to candidate events surviving the selection criteria; $n_{\rm S}$ and $n_{\rm B}$ are free parameters in the fit.

     For the $\lamc \ra \pkpi$ channel, we extract the $\lamc$ signal yields by fitting the $M(\pkpi)$ distribution. The signal PDF is a sum of two Gaussian functions with a common mean, and the background PDF is a second-order polynomial. All parameters of $F_{\rm S}$ and $F_{\rm B}$ are floated. The fit result is shown in figure~\ref{lamc-data-invM-fit} (Left), along with the pull distribution.
     The fitted signal yield is $N_{\rm norm} = 1472190\pm5726$, where the uncertainty is statistical. From the MC simulation, the mass resolution for $\lamc \ra \pkpi$ is 8 MeV/$c^2$.

     For the $\lamc \ra \petap$ channel, we first check the $M(\ppipieta)$ distribution from normalized $\eta'$ sidebands, as shown in figure~\ref{lamc-data-invM-fit} (Right). The distribution from the normalized $\eta'$ sidebands is smoothly falling, indicating a negligible contribution from $\Lambda_c^+ \ra p\pi^+\pi^-\eta$ decays. We subsequently fit the $M(\petap)$ distribution to extract the $\lamc$ signal yield.
     A sum of a Gaussian function and a Crystal Ball (CB) function~\cite{cb} is used as the signal PDF, and a second-order polynomial as the background PDF. The Gaussian and CB functions are fixed to have a common mean. All other parameters are floated in the fit. The fit result, along with the pull distribution, is shown in figure~\ref{lamc-data-invM-fit} (Right). A clear $\lamc$ signal is observed in the $M(\petap)$ distribution. The fitted signal yield is $N_{\rm sig} = 294\pm52$, where the uncertainty is statistical. The mass resolution for $\lamc \ra \petap$ is 13 MeV/$c^2$ from the MC simulation, which is the half-width at half maximum.
     The statistical significance of the $\lamc$ signal is 6.3$\sigma$, calculated from the difference of the logarithmic likelihoods, $-2 \rm ln(\mathcal{L}_0/\mathcal{L}_{\rm max}) = 59.2$, where $\mathcal{L}_0$ and $\mathcal{L}_{\rm max}$ are the maximized likelihoods without and with a signal component, respectively~\cite{wilks}. The significance takes into account the difference in the number of degrees of freedom in the two fits ($\Delta ndf$=7).
     Since the largest systematic uncertainty is due to the fit, as described in section~\ref{sec:syserror}, alternative fits to the $M(\petap)$ spectrum under different fit conditions are performed and the $\lamc$ signal significance is larger than 5.4$\sigma$ in all cases.

\begin{figure*}[h!tbp]
             \begin{center}
             \includegraphics[width=0.5\textwidth]{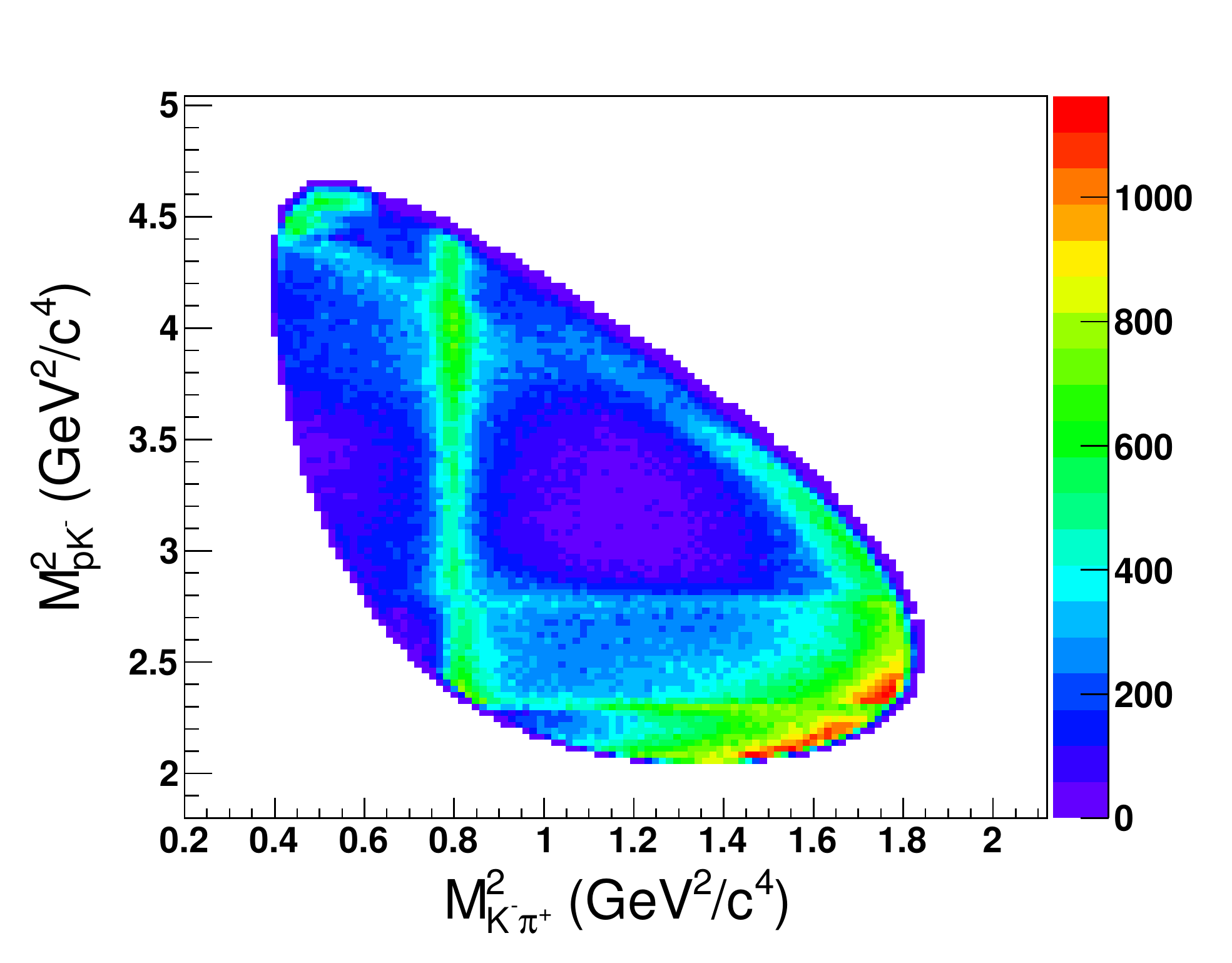}
             \caption{Dalitz plot of the selected $\lamc \ra \pkpi$ candidates.}
             \label{pkpi_data_dalitz}
             \end{center}
     \end{figure*}

     To measure the branching fraction, we must divide these extracted signal yields by their reconstruction efficiencies. Since $\lamc \ra \petap$ is a two-body decay and $\eta' \ra \eta\pi^+\pi^-$ is well modeled by the phase space~\cite{etapdecay}, we estimate the reconstruction efficiency directly from the simulated events by the ratio $n_{\rm sel}/n_{\rm gen}$, where $n_{\rm sel}$ and $n_{\rm gen}$ are the numbers of true signal events surviving the selection criteria and generated events, respectively. The signal mode reconstruction efficiency is determined to be $\epsilon_{\rm sig} = (2.22\pm 0.02)$\%. However, the reconstruction efficiency for the decay $\lamc \ra \pkpi$ can vary across the three-body phase space, as visualized in a Dalitz plot~\cite{dalitz} with polarization neglected. To take this into account, we correct the reconstruction efficiency according to the Dalitz plot from data as follows. Figure~\ref{pkpi_data_dalitz} shows the Dalitz distribution of $M^2(pK^-)$ versus $M^2(K^-\pi^+)$ in the $\lamc \ra \pkpi$ signal region from data which is defined as 2.274 $<$$M(\pkpi)$$<$ 2.298 GeV/$c^{2}$. The number of background events has been subtracted using the normalized $\lamc$ sidebands defined as (2.260, 2.272) GeV/$c^{2}$ and (2.300, 2.312) GeV/$c^{2}$. The effect of the variation of the kinematic boundaries with $M(\pkpi)$ is neglected. We divide the Dalitz plot for the data into 120$\times$120 bins, with a bin size of 0.027 $\rm GeV^{ 2}$$/c^{4}$ for $ M^{2}(pK^{-})$ and 0.016 $\rm GeV^{ 2}$$/c^{4}$ for $M^{2}(K^{-}\pi^{+})$.
     The corrected reconstruction efficiency is determined via the formula
     \begin{equation}
           \label{eq:pkpi-eff}
           \epsilon_{\rm norm}^{\rm corr} = \frac{\Sigma_{i}s_{i}}{\Sigma_{j}(s_{j}/\epsilon_{j})},
     \end{equation}
     where $i$ and $j$ run over all bins; $\Sigma_{i}s_{i}$ is the number of signal candidates in data; $s_{j}$ and $\epsilon_{j}$ are the number of signal candidates in data and the reconstruction efficiency from MC simulation events for each bin, respectively. The reconstruction efficiency for each bin is obtained by dividing the number of signal events after applying the selection criteria by the number of generated events. The corrected reconstruction efficiency is $\epsilon_{\rm norm}^{\rm corr} = (14.06\pm 0.01)$\%.

     The branching fraction of the signal mode relative to that of the normalization mode is determined via
     \begin{equation}
           \label{eq:bfratio-corr}
           \frac{\BF(\lamc \ra \petap)}{\BF(\lamc \ra \pkpi)} = \frac{N_{\rm sig}/\epsilon_{\rm sig}}{N_{\rm norm}/\epsilon_{\rm norm}^{\rm corr}} \times \frac{1}{\BF'},
     \end{equation}
     where $\BF'=\BF(\eta' \ra \eta\pi^+\pi^-) \times \BF(\eta \ra \GG)$~\cite{pdg}.
     Inserting all extracted signal yields and reconstruction efficiencies into eq.~\eqref{eq:bfratio-corr} gives the branching fraction ratio
     \begin{equation}
           \label{eq:bfratio-result}
           \frac{\BF(\lamc \ra \petap)}{\BF(\lamc \ra \pkpi)} = (7.54 \pm 1.32 \pm 0.73) \times 10^{-3},
     \end{equation}
     where the first error is statistical and the second is systematic, as evaluated in section~\ref{sec:syserror}. Multiplying both sides of eq.~\eqref{eq:bfratio-result} by the world average value $\BF(\lamc \ra \pkpi)=(6.28\pm0.32)\times10^{-2}$~\cite{pdg} gives
     \begin{equation}
           \label{eq:bf-result}
           \BF(\lamc \ra \petap) = (4.73 \pm 0.82 \pm 0.46 \pm 0.24)\times 10^{-4},
     \end{equation}
     where the third uncertainty is from the uncertainty in $\BF(\lamc \ra \pkpi)$.

\section{Systematic Uncertainties}
\label{sec:syserror}
     Table~\ref{tab:syserror} summarizes the sources of systematic uncertainties in measuring the ratio of branching fractions. These uncertainties are assessed as follows.
     \begin{table}[tbp]
          \centering
          \caption{\label{tab:syserror} The sources of the relative systematic uncertainties (\%) in the measurement of the ratio of branching fractions. }
          \begin{tabular}{|c|c|} \hline
                         Source                                  & Systematic uncertainties (\%) \\ \hline
                         PID efficiency                          &3.5    \\
                         Photon efficiency                       &4.0    \\
                         $\eta'$ mass window                     &0.8    \\
                        Best-candidate selection                &1.2   \\
                         Normalization mode fit                  &1.8    \\
                         Signal mode fit                         &7.6    \\
                         $\BF'$                                  &1.3    \\
                      Signal MC sample size                   &0.7    \\ \hline
                         Total                                   &9.7    \\ \hline
          \end{tabular}
     \end{table}

     \begin{itemize}
     	\item The momentum and angular distributions of protons from $\lamc \ra \pkpi$ and $\lamc \ra \petap$ are similar. Based on the study of inclusive $\Lambda$ decay $\Lambda \ra \petap$, the uncertainty from PID efficiency of the proton is below 0.1\% and is therefore neglected.
     	For the pion with the same charge in the numerator and denominator of eq.~\eqref{eq:bfratio-result}, the uncertainties in PID efficiencies only partially cancel because of differences in momentum distributions; Using a control sample of $D^{*+} \ra D^{0}\pi^+ $ with $D^0 \ra K^-\pi^+$ decay, we assign an uncertainty of 0.7\%.
     	Uncertainties of $1.6$\% and $1.2$\% are assigned for the $K$ and remaining $\pi$ identification efficiencies, respectively. The total systematic uncertainty from PID is 3.5\%, adding individual PID uncertainties linearly as they are highly correlated.

        \item Based on a study of radiative Bhabha events, a systematic uncertainty of 2.0\% is assigned to the photon efficiency for each photon, and the total systematic uncertainty from photon reconstruction is thus 4.0\%.

        \item The efficiency of the requirement on $M(\eta\pi^+\pi^-)$ is 96.3\% for data and 97.1\% for MC simulation, with a ratio of 99.2\%. We take the relative difference on efficiencies between data and MC simulation as the uncertainty due to the $\eta'$ mass window, which is 0.8\%.

        \item The uncertainty from the best-candidate selection is estimated by turning off the multiple candidate rejection and the difference in branching fraction ratio without best-candidate selection relative to the nominal one is taken as the uncertainty, which is 1.2\%.

        \item Uncertainties from the fits are estimated by modifying the forms of $F_{\rm S}$ and $F_{\rm B}$, and enlarging the fit range.
        \begin{itemize}
        	\item[$\bullet$] To evaluate the uncertainty from the fit of the normalization mode, the signal PDF used is a sum of three Gaussian functions with a common mean, and the background PDF used is a first-order polynomial. The fit range is enlarged to be (2.15, 2.42) GeV/$c^2$. The difference in signal yields is taken as the uncertainty. We sum these contributions in quadrature to obtain a 1.8\% uncertainty.

        	\item[$\bullet$] To evaluate the uncertainty from the fit of the signal mode, the signal PDF is changed to a sum of two CB functions, and the background PDF is changed to a third-order polynomial. We also enlarge the fit range. Moreover, we perform a simultaneous fit to the $M(p\pi^+\pi^-\eta)$ distributions from the selected events in the $\eta'$ signal region and the normalized $\eta'$ sidebands to estimate the contribution from a possible $\Lambda_c^+ \ra p\pi^+\pi^-\eta$ non-resonant component.
            The difference in signal yields from each change is taken as the uncertainty. We sum these contributions in quadrature to obtain a 7.6\% uncertainty.

        \end{itemize}

        \item The uncertainty on the product of the branching fractions of the decays $\eta' \ra \eta\pi^+\pi^-$ and $\eta \ra \GG$ is 1.3\%~\cite{pdg}.

        \item Uncertainty from the  finite statistics of the signal MC samples is 0.7\%.

     	\item Since there are three charged tracks in the final states both for normalization mode and signal mode, the uncertainty due to the tracking efficiency largely cancels in the ratio. For the normalization mode, the reconstruction efficiency is corrected by the Dalitz plot of the data; the uncertainty from the generator model is therefore negligible. For the signal mode, since the signal efficiency does not depend on the angular distribution of the proton in the $\lamc$ rest frame, the model-dependent uncertainty has a negligible effect on the efficiency.  For the decay $\eta' \ra \eta\pi^+\pi^-$, the internal dynamics is very similar to the phase space~\cite{etapdecay}, thus the systematic uncertainty from the generator model of $\eta' \ra \eta\pi^+\pi^-$ is ignored.
     \end{itemize}

     The total systematic uncertainty is obtained by adding in quadrature all the above contributions. The results are listed in table~\ref{tab:syserror}.

\section{Conclusions}
\label{sec:conc}
     In summary, using the full Belle data set corresponding to an integrated luminosity of 981 $\rm fb^{-1}$, we present the first measurement of the branching fraction of $\lamc \ra p\eta'$ relative to that of $\lamc \ra \pkpi$. The decay $\lamc \ra \petap$ is observed with a significance of 5.4$\sigma$. Our result is
     \begin{equation*}
     \frac{\BF(\lamc \ra \petap)}{\BF(\lamc \ra \pkpi)} = (7.54 \pm 1.32 (\rm stat) \pm 0.73 (\rm syst)) \times 10^{-3}.
     \end{equation*}
      Inserting the world average value $\BF(\lamc \ra \pkpi) = (6.28\pm0.32)\times10^{-2}$~\cite{pdg} gives the absolute branching fraction
     \begin{equation*}
     \BF(\lamc \ra \petap) = (4.73 \pm 0.82 (\rm stat) \pm 0.46 (\rm syst) \pm 0.24 (\rm ref))\times 10^{-4},
     \end{equation*}
     where the uncertainties are statistical, systematic, and from $\BF(\lamc \ra \pkpi)$, respectively.
     Our result is consistent with theoretical calculations based on $SU(3)_F$ symmetry~\cite{lamcR5, Ta3}, while not consistent with a constituent quark model prediction~\cite{lamcR3}.
\appendix

\acknowledgments
We thank the KEKB group for the excellent operation of the
accelerator; the KEK cryogenics group for the efficient
operation of the solenoid; and the KEK computer group, and the Pacific Northwest National
Laboratory (PNNL) Environmental Molecular Sciences Laboratory (EMSL)
computing group for strong computing support; and the National
Institute of Informatics, and Science Information NETwork 5 (SINET5) for
valuable network support.  We acknowledge support from
the Ministry of Education, Culture, Sports, Science, and
Technology (MEXT) of Japan, the Japan Society for the
Promotion of Science (JSPS), and the Tau-Lepton Physics
Research Center of Nagoya University;
the Australian Research Council including grants
DP180102629, 
DP170102389, 
DP170102204, 
DP150103061, 
FT130100303; 
Austrian Federal Ministry of Education, Science and Research (FWF) and
FWF Austrian Science Fund No.~P~31361-N36;
the National Natural Science Foundation of China under Contracts
No.~11475187,  
No.~11521505,  
No.~11575017,  
No.~11675166,  
No.~11705209,  
No.~12005040,
No.~11761141009,
No.~11975076,
No.~12042509,
No.~12135005,
No.~12161141008;
Key Research Program of Frontier Sciences, Chinese Academy of Sciences (CAS), Grant No.~QYZDJ-SSW-SLH011; 
the  CAS Center for Excellence in Particle Physics (CCEPP); 
the Shanghai Science and Technology Committee (STCSM) under Grant No.~19ZR1403000; 
the Ministry of Education, Youth and Sports of the Czech
Republic under Contract No.~LTT17020;
Horizon 2020 ERC Advanced Grant No.~884719 and ERC Starting Grant No.~947006 ``InterLeptons'' (European Union);
the Carl Zeiss Foundation, the Deutsche Forschungsgemeinschaft, the
Excellence Cluster Universe, and the VolkswagenStiftung;
the Department of Atomic Energy (Project Identification No. RTI 4002) and the Department of Science and Technology of India;
the Istituto Nazionale di Fisica Nucleare of Italy;
National Research Foundation (NRF) of Korea Grant
Nos.~2016R1\-D1A1B\-01010135, 2016R1\-D1A1B\-02012900, 2018R1\-A2B\-3003643,
2018R1\-A6A1A\-06024970, 2019K1\-A3A7A\-09033840,
2019R1\-I1A3A\-01058933, 2021R1\-A6A1A\-03043957,
2021R1\-F1A\-1060423, 2021R1\-F1A\-1064008;
Radiation Science Research Institute, Foreign Large-size Research Facility Application Supporting project, the Global Science Experimental Data Hub Center of the Korea Institute of Science and Technology Information and KREONET/GLORIAD;
the Polish Ministry of Science and Higher Education and
the National Science Center;
the Ministry of Science and Higher Education of the Russian Federation, Agreement 14.W03.31.0026, 
and the HSE University Basic Research Program, Moscow; 
University of Tabuk research grants
S-1440-0321, S-0256-1438, and S-0280-1439 (Saudi Arabia);
the Slovenian Research Agency Grant Nos. J1-9124 and P1-0135;
Ikerbasque, Basque Foundation for Science, Spain;
the Swiss National Science Foundation;
the Ministry of Education and the Ministry of Science and Technology of Taiwan;
and the United States Department of Energy and the National Science Foundation.

\end{document}